\documentclass[sigconf]{acmart}
\usepackage{listings}
\usepackage{float}
\AtBeginDocument{%
  }

\copyrightyear{2025}
\acmYear{2025}
\setcopyright{cc}
\setcctype{by}
\acmConference[IMC '25]{Proceedings of the 2025 ACM Internet Measurement Conference}{October 28--31, 2025}{Madison, WI, USA}
\acmBooktitle{Proceedings of the 2025 ACM Internet Measurement Conference (IMC '25), October 28--31, 2025, Madison, WI, USA}
\acmDOI{10.1145/3730567.3764492}
\acmISBN{979-8-4007-1860-1/2025/10}

\begin{document}

\title[On YouTube Search API Use in Research]{On YouTube Search API Use in Research}

\author{Alexandros Efstratiou}
\affiliation{%
    \institution{University of Washington}
    \city{Seattle}
    \state{WA}
    \country{USA}
}

% %%
% %% By default, the full list of authors will be used in the page
% %% headers. Often, this list is too long, and will overlap
% %% other information printed in the page headers. This command allows
% %% the author to define a more concise list
% %% of authors' names for this purpose.

\begin{abstract}
    YouTube is among the most widely-used platforms worldwide, and has seen a lot of recent academic attention.
    Despite its popularity and the number of studies conducted on it, much less is understood about the way in which YouTube's Data API, and especially the Search endpoint, operates.
    In this paper, we analyze the API's behavior by running identical queries across a period of 12 weeks.
    Our findings show that the search endpoint returns highly variable results between queries.
    Specifically, the API seems to randomize returned videos based on the relative popularity of the respective topic during the query period, making it nearly impossible to obtain representative historical video samples, especially during non-peak topical periods.
    Our results also suggest that the API may prioritize shorter, more popular videos, although the role of channel popularity is not as clear.
    We conclude with suggested strategies for researchers using the API for data collection, as well as future research directions on expanding the API's use-cases.
\end{abstract}




% \received{15 May 2025}

\maketitle

\section{Introduction}\label{sec:intro}

YouTube has been used to study a wide array of crucial societal problems like online hate~\cite{albadi_deradicalizing_2022,papadamou_how_2021}, accessibility~\cite{may_unspoken_2024}, pseudoscientific misinformation~\cite{papadamou_it_2022}, online scams~\cite{liu_give_2024}, and child exposure to inappropriate content~\cite{gkolemi_youtubers_2022,khan_analyzing_2024,papadamou_disturbed_2020}.
Much of this work has made use of the YouTube Data API,
%\footnote{\url{https://developers.google.com/youtube/v3/docs}} 
which offers several endpoints.
%to retrieve video metadata and comments by video ID, return search results using keywords, and more.
However, researchers have pointed out that the YouTube API may sometimes return seemingly biased results~\cite{choi_creator_2025,violot_shorts_2024}.
Beyond documentation and implementation errors~\cite{martin-lopez_restest_2021}, some works suggest that this can result from systematic API behavior, particularly through the keyword-based Search: list endpoint (``search'' henceforth), which makes obtaining random samples difficult~\cite{mcgrady_dialing_2023,rieder_ranking_2018}.

Although YouTube data are collected in a variety of ways, including crawling through recommended videos~\cite{mamie_are_2021,ribeiro_auditing_2020,rochert_homogeneity_2020}, deploying sockpuppet accounts~\cite{hussein_measuring_2020,jung_algorithmic_2025}, extensions installed on participants' browsers~\cite{hu_u_2024}, or collecting videos from pre-curated channels~\cite{bertaglia_monetisation_2024,khan_analyzing_2024}, the search endpoint remains an important part of the data collection pipeline for a lot of research~\cite{albadi_deradicalizing_2022,may_unspoken_2024,nigatu_i_2024,papadamou_disturbed_2020,papadamou_how_2021,pera_shifting_2024,violot_shorts_2024}.
Given its importance and the fact that its behavior is poorly understood~\cite{choi_creator_2025,mcgrady_dialing_2023}, we conduct an audit of this API endpoint.
We run identical queries at 5-day intervals across a period of approximately 3 months to answer the following research questions:

\begin{itemize}
    \item[\textbf{RQ1}] How consistent are the data returned by the Search endpoint?
    \item[\textbf{RQ2}] How does the API determine which videos are returned?
\end{itemize}

Comparing the set similarities of videos obtained with each query, we find that this similarity decays over time, indicating that datasets collected using the exact same historical query may differ vastly based simply on when the queries were made.
Moreover, we analyze the historical time points on which the most data is returned.
We find that the API may be withholding data for periods of relative topical inactivity even though returning these data would not bring the response above the maximum number of videos allowed by the API.
Using a second-order Markov chain to model ``transitions'' between the presence or the absence of a video in successive collections, we find that video omission or inclusion is mostly conducted in a ``rolling window'' fashion.
Finally, we analyze whether any video metadata (e.g., likes, views, etc.) are associated with more consistent video returns, finding that the API is more likely to return more popular videos that are drawn from less active topics.

Through this work, we aim to inform better search strategies when using the YouTube API in terms of replicability and API token economy, while also exploring new ways in which the search endpoint can be used in academic research.

%\alex{check that all of these are met before submitting: \url{https://conferences.sigcomm.org/imc/2025/submission-instructions/}}
\section{Background}\label{sec:background}

%When discussing use of the search API in research, it is worth first understanding the officially documented limitations of the API, how other works have used it, and how researchers have noted unusual behaviors that are not documented.

%\paragraph{Overview of the Search: list endpoint.}

Based on the official documentation,\footnote{\url{https://developers.google.com/youtube/v3/docs/search/list}} the search endpoint allows a user to search by keywords, location, or live events, and enables filtering by several other parameters such as date ranges, specific channels, etc.
However, this endpoint has a quota cost of 100 units per query;
%(which includes pagination)
this is considerably higher than ID-based endpoints, which typically only cost 1 unit.
With the default daily quota being 10,000, this allows 100 search queries per day per client.
However, the YouTube Data API has a researcher access program that allows higher quotas to vetted accounts.
This endpoint is not designed for volume.
The maximum number of results per query is 500 (max. 50 per page and max. 10 pages)~\cite{yin_smappnyuyoutube-data-api_2018}, thus, it is necessary for YouTube to sample the videos it returns.

One approach to collect all videos on a topic was to identify topical ``seed'' channels and videos (either through external sources~\cite{papadamou_disturbed_2020} or keyword search~\cite{papadamou_it_2022}), and obtain videos recommended by YouTube as being relevant to those initial sets to expand the dataset.
However, the \texttt{relatedToVideoId} response that enabled this approach was deprecated in 2023, eliminating it from being conducted through the API.

An alternative advocated strategy has been time-split queries for clients endowed with sufficient quota~\cite{yin_smappnyuyoutube-data-api_2018}.
The API allows the client to add parameters for \texttt{publishedAfter} and \texttt{publishedBefore} to restrict the data collection period.
Researchers have used this to query in a ``one per X time'' fashion~\cite{rieder_ranking_2018}, where the observation period is split into time bins, each of which is queried separately to circumvent the 500-video limit~\cite{porreca_using_2020,violot_shorts_2024}.
In theory, this should enable researchers to obtain every video uploaded on the topic, unless more than 500 videos were uploaded on a specific day.
However, recent work finds a strong recency bias with this approach, with a much higher volume of videos for dates closer to the query date than for historical dates~\cite{rieder_forgetful_2025}.

Another commonly used strategy is identifying relevant channels through external sources like SocialBlade~\cite{khan_analyzing_2024} or Reddit~\cite{bertaglia_monetisation_2024} and querying the API for their videos.
While this can be done using several endpoints, for example, using Channels: list to extract a ``playlist'' of a channel's uploaded videos and then querying the PlaylistItems: list endpoint for those videos, or adding a ``channelId'' parameter to the search endpoint, few papers clarify the exact endpoints used~\cite{may_unspoken_2024} or the dates on which the queries themselves are made~\cite{papadamou_it_2022}.
As we later show, both of these can influence returned data.

\section{Methods}

For our experiments, we choose a diverse range of political, scientific, and entertainment topics that are either regional or international and vary in size and recency.
Although these topics are not exhaustive, they allow us to observe whether certain patterns may be due to characteristics like topic sensitivity.
The exact queries per topic are shown in Appendix~\ref{app:queries}:

\begin{itemize}
    \item Black Lives Matter (BLM) (2020)
    \item Brexit (2016)
    \item US Capitol Riots (2021)
    \item Grammy Awards (2024)
    \item Higgs Boson (2012)
    \item World Cup (2014)
\end{itemize}

Each topic has a focal ``D-day'' on which a central event took place (e.g., the day of the referendum for Brexit; see Appendix~\ref{app:queries}).
We set our data collection period between two weeks before and after this date (i.e., a total collection span of 28 days per topic).
For each topic, we run the same query every five days through the YouTube Data API v3 Search: list endpoint, starting on February 9 and ending on April 30, 2025.
Due to a technical problem, the collection on April 5th was skipped.
Our data and code are available on GitHub.\footnote{\url{https://github.com/alefstrat/youtube_api_audit}}
% (hence the gap in some of the figures henceforth).

We send queries for every hour within these 28 days to circumvent the maximum limits in returned videos imposed by the API following the strategy outlined in Section~\ref{sec:background}, resulting in 4032 total queries for every collection (24 hours $\times$ 28 days $\times$ 6 topics) and 16 snapshots over 12 weeks.
Queries are made with an API token, not OAuth 2.0, so that account effects do not confound our results.
Responses are set to be returned in reverse chronological order.
We choose this order as it is an immutable video property, whereas other ordering options like view count or relevance may change over time.
Thus, it offers the best baseline to study API consistency.
However, we stress that the YouTube API’s documentation makes no commitment to order results beyond a daily granularity.
Our choice to make queries for each individual hour is so that we maintain more consistency between low-activity and high-activity days, i.e., avoiding ceiling effects for days on which more than 500 videos may have been posted.
However, on such days, it can alternatively be the case that order may take precedence over time-filtering. 
That is, returned videos may be sampled from a given day, rather than a given hour, if the number of eligible results exceeds the maximum.

\section{API Behavior}

In this section, we highlight how the API's search endpoint varies returned results.
We document some of the potential mechanisms of this variability and how videos may be drawn when determining what to return to the client.

\subsection{Temporal Variability}

\begin{figure*}[t!]
  \centering
    \includegraphics[width=0.75\textwidth]{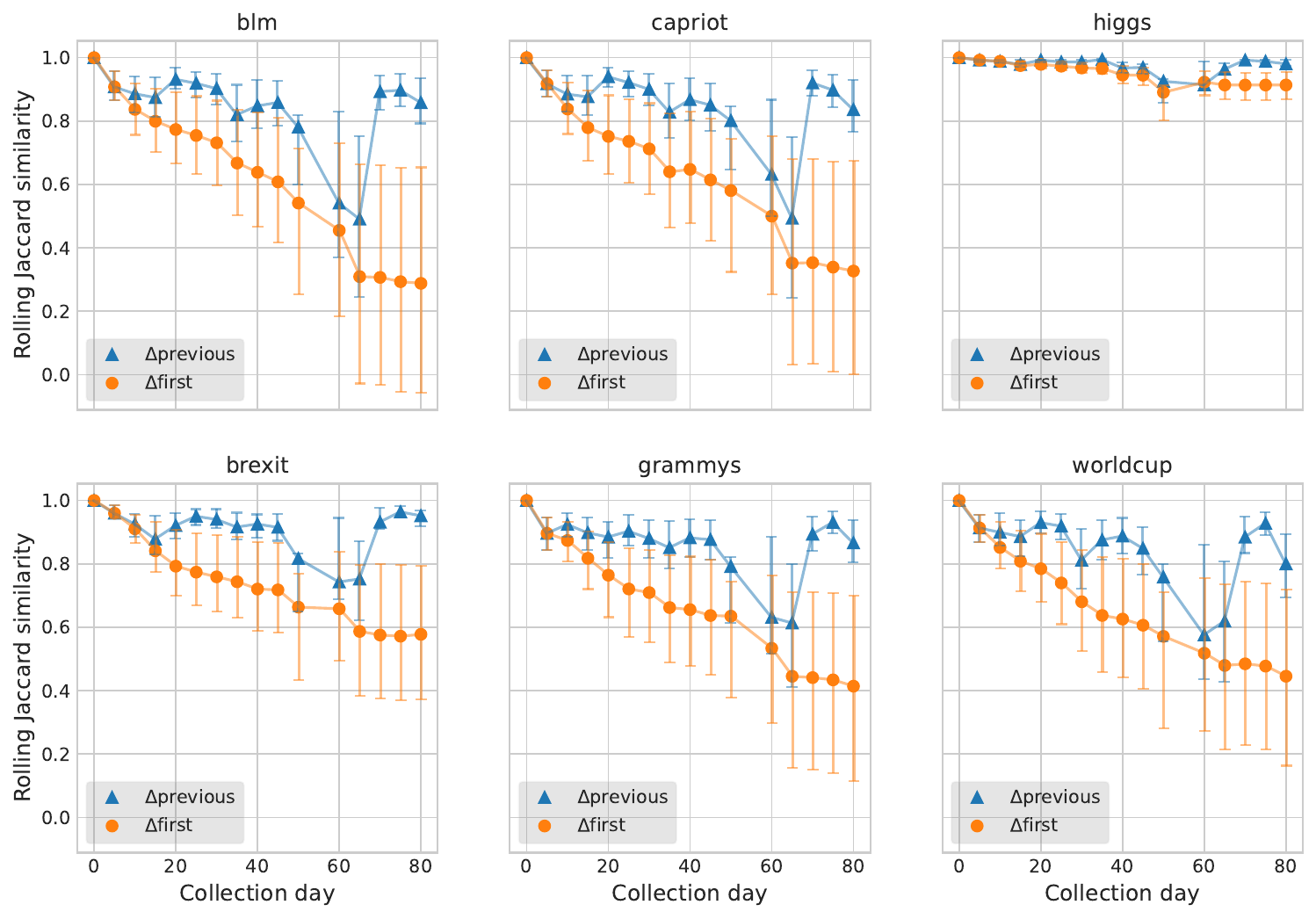}
    \caption{Jaccard similarities of video ID sets relative to the previous and the first collection instance using the ``search'' endpoint. ``Error bars'' represent the set difference of $S_{t-1} - S_t$ (bottom bars) and $S_t - S_{t-1}$ (top bars).}
    \label{fig:vid_jac}
\end{figure*}

For every collection instance at time $t$, we obtain the set of video IDs returned $S_t$ and calculate its Jaccard similarity with the set obtained in the previous collection $S_{t-1}$ and the very first collection $S_{t-n}$.
We plot these rolling Jaccard similarities in Figure~\ref{fig:vid_jac}.
Table~\ref{tab:numvids} shows descriptives for the total number of videos returned per collection.

\begin{table}[t!]
    \small
    \centering
    \begin{tabular}{lrrrr}
    \toprule
        \textbf{topic} & \textbf{min} & \textbf{max} & \textbf{mean} & \textbf{std} \\
        \midrule
        BLM & 639 & 765 & 743.44 & 27.86 \\
        Brexit & 478 & 573 & 559.81 & 21.86 \\
        Capitol & 507 & 590 & 571.81 & 17.35 \\
        Grammys & 564 & 677 & 659.13 & 25.45 \\
        Higgs & 476 & 512 & 507.44 & 8.32 \\
        World Cup & 419 & 516 & 502.5 & 21.96 \\
        \bottomrule
    \end{tabular}
    \caption{Descriptive statistics for number of videos returned per topic across collections.}
    \label{tab:numvids}
\end{table}

We find differences between successive runs, which, over time, compound to form video sets that are vastly different from the initial collection.
% , to Jaccard values as low as $\sim$0.3 after 3 months.
% This equates to only 46\% of the videos per set being shared, with the majority being unshared.
The exception is the Higgs topic, which retains much higher consistency than the rest--we offer an explanation of why that may be in Section~\ref{sec:regressions}.
The ``error bars'' in Figure~\ref{fig:vid_jac} rule out content deletions as a potential explanation;
% As shown by the ``error bars'' in Figure~\ref{fig:vid_jac}, content deletions cannot explain this phenomenon as 
we find videos at $t$ that were not seen at $t-x$, despite these queries being fully historical (i.e., not spanning the query date).
This behavior does not extend to endpoints that take IDs as queries (see Appendix~\ref{app:id_endpoints}).
% Moreover, we verify that this behavior does not extend to other endpoints that use ID-based queries (see Appendix~\ref{app:id_endpoints}).

\subsection{Randomization Mechanisms}

An obvious question is whether the variability is due to ceiling effects, i.e., whether randomization occurs due to the possible matches exceeding the maximum number of results allowed to be returned.
% despite the query specifying that results should be returned in chronological order.
% That is, if the eligible videos for a particular query exceed the maximum of 50 per page/500 per query allowed by the API, then we might expect that the results returned will be randomized.
% In that case, we should expect that some of our queries return more than 500 eligible videos, and that Jaccard similarities between collections are lower for hours on which more videos are uploaded.
In this section, we test this theory.

We first obtain some descriptive statistics of videos returned for each hour and each of the topics (Table~\ref{tab:descs}).
The maximum number of videos returned for any given hour (max) remains well below the theoretical maximum of 50 per page, ruling out the ceiling effect explanation.
Moreover, we compute the Spearman coefficient $\rho$ for the correlation between the Jaccard similarity of sets $T_1$ (first collection) and $T_L$ (last collection) and the average number of videos returned for that hour.
However, to avoid inflating Jaccard similarity values based on empty sets, we first drop all hours for which 0 videos are returned across collections.
The correlation is meant as a soft test of the ceiling effect, since, if randomization is indeed more prominent when more videos are returned, we should expect lower Jaccard similarities with a higher number of returned videos (i.e., a negative correlation).
However, we observe almost the exact opposite pattern: For all but the Higgs topic, for which we observe a non-significant (negative) correlation, there are weak positive correlations between the number of videos for that hour and the Jaccard similarity, indicating that similarity values are, on average, higher for busier hours.
Although this may be an artifact of the idea that more videos simply stabilize the Jaccard value, this analysis demonstrates that fewer videos do not necessarily mean that those videos will be the same across collections.

\begin{table}[t!]
    \small
    \centering
    \begin{tabular}{lrrrrrr}
    \toprule
        \textbf{topic} & \textbf{mean} & \textbf{min} & \textbf{max} & \textbf{std} & $\rho$ & \textbf{\textit{N}} \\
        \midrule
        BLM & 1.10 & 0 & 17 & 2.33 & **0.13 & 267 \\
        Brexit & 0.83 & 0 & 13 & 1.57 & ***0.15 & 324 \\
        Capitol & 0.85 & 0 & 28 & 2.54 & ***0.29 & 242 \\
        Grammys & 0.98 & 0 & 21 & 2.22 & ***0.26 & 387 \\
        Higgs & 0.75 & 0 & 14 & 1.62 & -0.11 & 216 \\
        World Cup & 0.75 & 0 & 31 & 1.37 & *0.12 & 418 \\
        \bottomrule
    \end{tabular}
    \caption{Descriptive statistics for per-hour number of videos returned. *p < 0.05, **p < 0.01, ***p < 0.001. \textit{N} is the number of videos retained after all hours with no videos returned across collections are dropped.}
    \label{tab:descs}
\end{table}

\begin{figure*}[t!]
  \centering
    \includegraphics[width=0.99\textwidth]{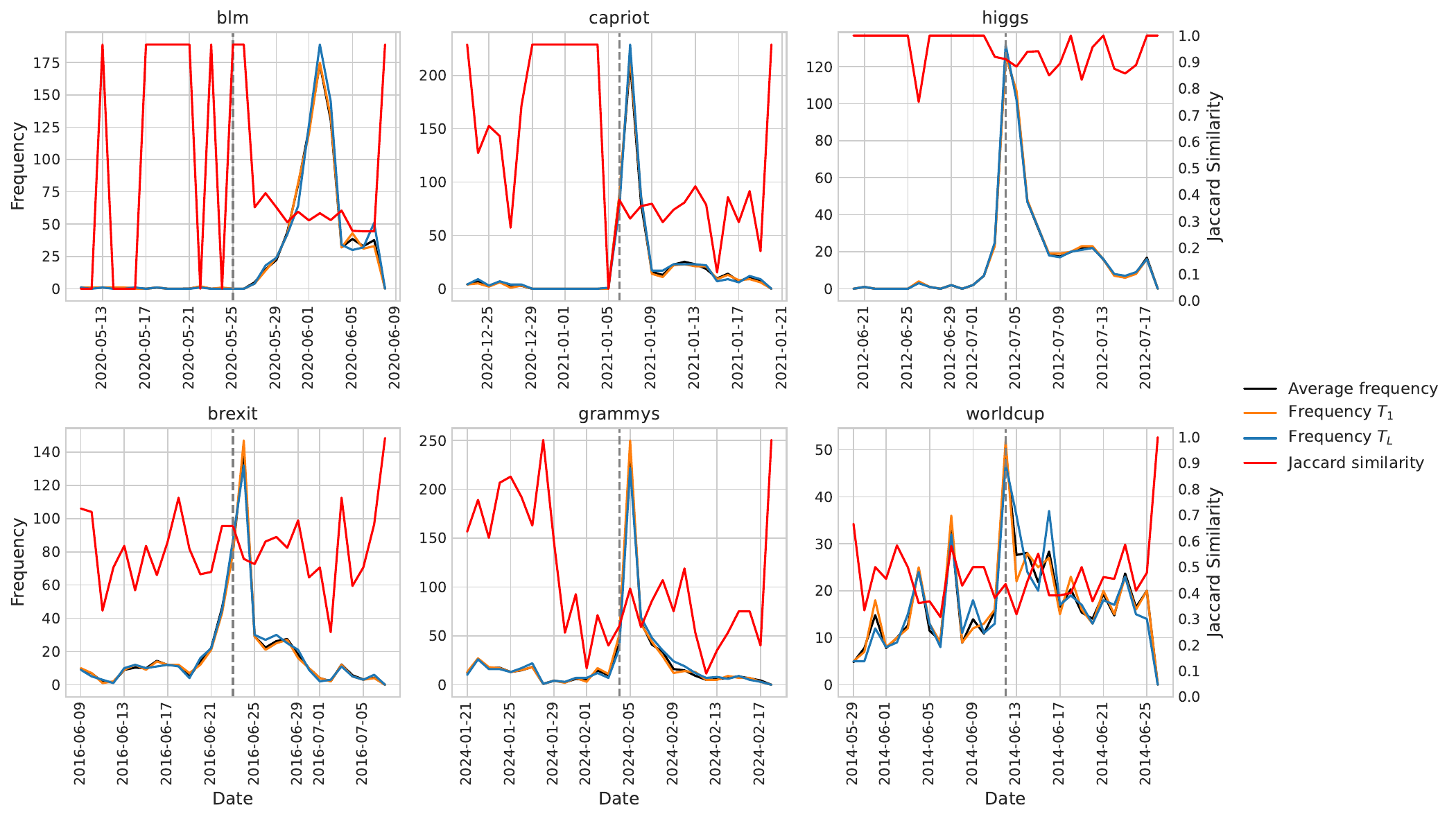}
    \caption{Daily frequencies of videos returned with daily Jaccard similarities between first and last collections. Dashed vertical lines represent the corresponding topic's D-day.}
    \label{fig:daily_freqs}
\end{figure*}

% To better understand how video volumes are distributed over time, 
We also plot the daily video frequencies of the first and last collections alongside the Jaccard similarity for these daily sets in Figure~\ref{fig:daily_freqs}.
To ensure that these are not edge cases, we further plot the average daily frequencies across all collections.
As can be seen, the average daily frequency distributions per collection map almost perfectly on each other.
However, the volume of videos returned does not map onto the Jaccard similarities in any consistent manner, confirming our above findings of weak or non-existent correlations between these two factors.
Most videos returned are uploaded around the focal date, with the exception of the BLM topic (likely due to when the protests surrounding George Floyd's death intensified; the topical peak is recorded on Blackout Tuesday).
Jaccard similarities at frequency peaks are comparable to other days on which much fewer videos are uploaded, indicating that the YouTube Data API operates on time-dependent systematic randomization.
Interestingly, the YouTube recommendation system uses an empirical distribution of video popularity against its age to determine how relevant it is to recommend to users~\cite{covington_deep_2016}.
These patterns suggest that the API may similarly sample from topic-wide empirical distributions in terms of over-time interest.

Overall, we do not find evidence of ceiling effects driving the API randomization.
Instead, our results suggest that the YouTube API samples videos from empirical distributions, returning results based on the relative density of topical interest and even forcing zero videos to be returned when this relative density is adequately low.
However, it is unclear how this interest is computed (e.g., if it is the volume of videos uploaded or something else).
% As such, the randomization process seems to be representative of empirical video upload frequencies over time, even forcing zero videos to be returned when this relative density is adequately low.
% We make this claim under the assumption that each of the hourly queries we make are \textit{truly} independent of each other (i.e., no token-based interdependencies), and so they should, in theory, return independent results.
% (which does not seem to be the case).

\subsection{Attrition Analysis}

\begin{figure}[t!]
  \centering
    \includegraphics[width=0.85\columnwidth]{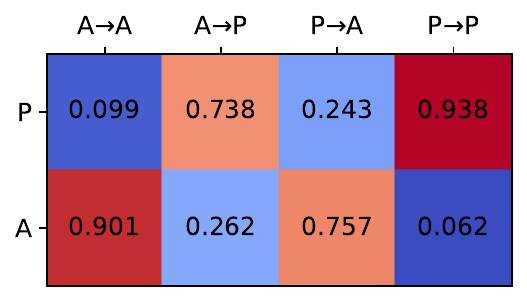}
    \caption{Transition probabilities for presence (P) or absence (A) of videos in a second-order Markov chain. Example interpretation: Column $A\rightarrow A$ and row $P$ show the probability of P given that the two previous states are A.}
    \label{fig:trans_prob}
\end{figure}

Next, we focus on whether this sampling operates on a ``drop-in/drop-out'' basis; that is, are videos more likely to reappear or remain left out in successive collections?
To answer this, we utilize a second-order Markov chain where we treat the presence (P) or absence (A) of a video in any given collection as the two possible states.
Then, across all topics and videos, we compute the transition probability from the two most recent states to the next one in a sliding window.
We show the resulting transition probabilities in Figure~\ref{fig:trans_prob}.

The results suggest that drop-ins and drop-outs are the normative behavior.
That is, a video is more likely to be present or absent in a collection if it is present or absent, respectively, in the immediately previous collection.
Moreover, this probability is higher when both previous states are the same.
Therefore, the probability that a video is returned in the collection set may be influenced not only by the video's upload date, but by the request date itself and whether the video is in the ``windowed set'' for that date.

\section{Factors Behind Return Likelihood}\label{sec:regressions}

Our next analysis focuses on whether the YouTube API is more likely to return videos with certain features at higher rates.
We treat this as an explainability problem where we count the number of times each video is returned in our API calls.
We then determine the features that are most predictive of this frequency.
Our candidate predictors are topic, video duration, video definition, number of views, comments, and likes at the video level; channel views, subscribers, and number of videos uploaded at the channel level.
All continuous features are log-transformed to normalize their distributions.

\begin{figure}[t!]
    \centering
    \includegraphics[width=0.99\columnwidth]{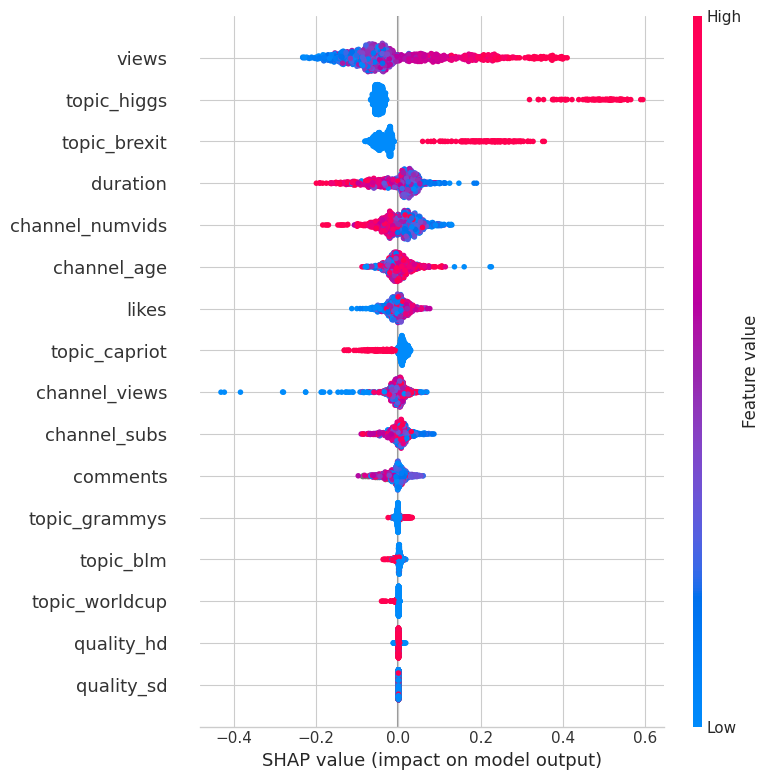}
    \caption{Beeswarm plot of SHAP values per feature.}
    \label{fig:shap_beeswarm}
\end{figure}

We deploy a gradient boosting model using LightGBM~\cite{ke_lightgbm_2017} with a Poisson regression objective, 100 estimators, and a learning rate of 0.1.
% (default parameters other than the objective).
The data are split into 80-20 train-test sets ($R^2 = 0.19)$.
We examine the role of each feature from the predictions on the test set using SHapley Additive exPlanations (SHAP)~\cite{lundberg_unified_2017}, shown in Figure~\ref{fig:shap_beeswarm}.
Features are shown in descending order of importance.
We also perform robustness checks using regression models with the same variables in Appendix~\ref{app:regressions}, which show directionally consistent results.

Starting with video metadata, we see that the number of views tends to positively predict return frequency and is the most important feature.
In terms of duration, which is also among the most important features, shorter videos tend to have more returns.
There is also a slightly positive effect of likes, while the number of comments is not as important and video quality plays no role.
Overall, this suggests that the API tends to return shorter, more popular videos.

For channel features, returns are more likely from older channels with fewer videos.
% ; these features are similarly important.
There is a slightly negative trend in terms of the number of channel subscriptions, while the role of accumulated number of channel views is more mixed.
Overall, channel popularity is not as important, though other channel features like age and activity may play a role.

Interestingly, the Higgs Boson and Brexit topics are highly important and positive features for return frequency; the Grammy's topic also trends towards positive effects.
%, but is not as important.
The Capitol riots, BLM, and World Cup topics trend toward negative effects, though the latter two are not as important.
% as the first.

Although the directionally grouped topics look unrelated at first, a deeper look reveals that they are linked by one factor: size.
We determine this using metadata returned with every query we send, which contain a value called \texttt{pageInfo.totalResults}.
This reflects the total number of results in the result set,
% \footnote{\url{https://developers.google.com/youtube/v3/docs/search/list}}
i.e., the total number of videos that match our query, with a maximum potential value of 1,000,000.
We obtain the minimum, maximum, mean, and modal values from this field across every hour and collection run that we query per topic.
As seen in Table~\ref{tab:topic_sizes}, the three topics whose videos have higher appearance frequencies are also the smallest (and the only ones without a modal value of 1M, which is the maximum).
The results indicate that queries with a smaller video pool to draw from may return more consistent results (and may explain why Higgs, which is by far the smallest topic, is also the most consistent in Figure~\ref{fig:vid_jac}), though this remains to be experimentally verified.

\begin{table}[t!]
    \small
    \centering
    \begin{tabular}{lrrrr}
    \toprule
        \textbf{Topic} & \textbf{Min} & \textbf{Max} & \textbf{Mean} & \textbf{Mode} \\
        \midrule
        BLM & 679k & 1M & 982k & 1M \\
        Brexit & 247k & 786k & 624k & 613k \\
        Capitol Riot & 515k & 1M & 966k & 1M \\
        Grammys & 12.8k & 1M & 150k & 123k \\
        Higgs & 5.50k & 65.2k & 40.2k & 39.0k \\
        World Cup & 634k & 1M & 998k & 1M \\
        \bottomrule
    \end{tabular}
    \caption{Potential video pool size per topic.}
    \label{tab:topic_sizes}
\end{table}

Notably, despite the modal number of videos \textit{returned} across all collection hours for all topics being 0, the modal value for the available \textit{pool} is much higher (and the maximum of 1M for 3/6 topics).
Since it is unlikely that 1M videos are consistently uploaded every hour for any of these topics, 
%Given the very low likelihood that 1M videos are consistently uploaded every hour for any of these topics and the fact that this pool number changes across collection runs,
this suggests that time constraints in queries do not affect the total pool of available videos in the API's results set.
Rather, they may only \textit{filter} the response \textit{after} results are returned.
% the API does not take into account time constraints in determining the total pool of available videos.
% This may be a mechanism via which the number of returned results is further restricted.
% Although this sounds counter-intuitive, recent work (\alex{CITE Rieder et al., in press}) has demonstrated that the YouTube Search API has a severe recency bias such that it returns substantially more videos posted within approximately 3 weeks of the \textit{query's} date.
% Thus, the API may be fixing higher probability towards these videos, which would inherently mean that older videos have a lower probability of being picked out of the pool as they make up a much smaller proportion of the total.

Comparing Tables~\ref{tab:numvids} and ~\ref{tab:topic_sizes}, it is also striking that the number of videos \textit{returned} is much closer across topics than their respective topic sizes suggest, which is consistent with our distribution density explanation.
This becomes even more apparent when scrutinizing the y-axis on Figure~\ref{fig:daily_freqs}: The most-populated peaks are recorded for topics where the rest of the time-series is relatively inactive (e.g., Capitol Riot, Grammys), while topics that are more active throughout (e.g., World Cup, which is an ongoing tournament rather than a one-off event) record peaks at lower absolute values, most likely due to the \textit{number} of videos to be returned being fixed and the actual videos being drawn from an empirical distribution.

% \alex{explain this much better; basically, smaller topics, not a lot of recency bias because the density is still higher for older videos. newer topics that are still active: tough luck}
% \alex{For a breakdown of \texttt{pageInfo.totalResults} values per collection instance and observation period date, we refer the reader to Appendix X}.
% \alex{Figure out what we actually want to do with this: try and untangle the pattern a bit. if we want to expand the above discussion (and we might want to because it's pretty important) we might want to instead include those plots in the main body}
% \alex{Explain also the grammys findings, we ran the query while grammys was ongoing for the first few collections and so this might explain the sharp drop afterwards, consistent with our explanation that the API may be artificially inflating or deflating the available pool...}

\section{Discussion}

In this short paper, we contribute to the understanding of a crucial research tool.
The behaviors we record offer new strategies of working with the YouTube API and future research directions on expanding its use-cases.

\subsection{Implications}

Our findings show that binning queries across the observation period is not as fruitful as previously thought~\cite{rieder_ranking_2018,violot_shorts_2024,yin_smappnyuyoutube-data-api_2018}, and offers low return on investment considering the quota cost of the search endpoint.
Instead, researchers may experiment with breaking up their \textit{topics} as opposed to their time frames.
This can be achieved by incorporating more AND statements in the query or querying for multiple sub-topics (e.g., specific players alongside their national teams instead of the entirety of the World Cup event).
The total number of results in the query metadata is a crucial way of assessing how optimal a query is (with lower being better/more stable).
To reduce this number of results as much as possible, researchers should build their queries around API parameters that are specified in the documentation to affect \textit{search results} (e.g., query, region code, etc.), and not merely the \textit{API response} (e.g., before-after datetimes, topic IDs, etc.)

Alternatively, in cases where discovery is possible at the channel level instead of using keywords, or where consistency takes precedence over data completeness, ID-based endpoints are a viable route.
For example, as we outline in Section~\ref{sec:background}, a combination of the Channels: list and PlaylistItems: list endpoints, both of which take IDs as queries, would allow for the retrieval of complete channel uploads.
We also urge researchers to specify these endpoints in their data collection pipelines, as these choices can massively impact the replicability of their work.

\subsection{Limitations and Future Work}

The strategy of using parameters that restrict the potential results pool is not explicitly quantified here, and can be addressed in future work that designs progressively restrictive queries using other parameters as we do here with time constraints.
% The new strategies we put forward may need experimental validation, for example, by running progressively more restrictive queries and seeing how that influences the replicability of the data returned (alongside the reported video pool size).
Moreover, we do not explicitly analyze how other parameters, such as ordering or including channel IDs, may affect replicability.
%, although we have no reason to believe that this would alter our conclusions (especially as reverse chronological order should be immutable for historical queries, other than video deletions).

Future work can also replicate our experiments with more sparse collections over a longer period, to check for potential periodicity in set similarities.
Moreover, given the substantial efforts that scholars have expended in creating sockpuppets for YouTube SERP audits~\cite{hussein_measuring_2020,jung_algorithmic_2025}, similar methods to ours can be employed to check the consistency between results of sockpuppet SERPs and search endpoint results.
This would help us understand if the search endpoint has research value beyond data collection, for example, as a low-resource way of conducting SERP audits.

\section*{Acknowledgments}

This work has been supported by the University of Washington’s Center for an Informed Public, the John S. and James L. Knight Foundation (G-2019-58788), and the William and Flora Hewlett Foundation (2023-02789).
The author is grateful to Adrian Padilla, Bernhard Rieder, Ethan Zuckerman, and the anonymous reviewers for their helpful feedback.

\bibliographystyle{ACM-Reference-Format}
\balance
\bibliography{references}

\appendix

\section{Ethics}

This work makes sole use of video, comment, or channel IDs and high-level metadata (number of views, likes, etc.) as data points and does not analyze content beyond this point.
The aim of this paper is to offer a better understanding of how data from YouTube, one of the largest platforms worldwide, should be understood and used by researchers. The societal benefits arising from this work apply insofar as the important topics studied on YouTube that we cover in the Introduction benefit from the directions we offer.

\section{Query Parameters}\label{app:queries}

Our general query parameters, followed by topic-specific parameters (keywords and dates).

\subsection{General parameters}

Unless [variable], these parameters were kept consistent across queries.
% Note that \texttt{maxResults} refers to the maximum number of results per page, not across all pages.

\begin{verbatim}
    {
        "part": "snippet",
        "maxResults": 50,
        "order": "date",
        "safeSearch": "none",
        "publishedAfter": [variable],
        "publishedBefore": [variable],
        "type": "video",
        "q": [variable]
    }
\end{verbatim}

\subsection{Topic-specific parameters}

Keywords (q) and dates queried.
Note that, for dates, we passed ``publishedAfter'' as the topic-specific date -14 and ``publishedBefore'' as +14 days.

\paragraph{Black Lives Matter}

Focal date: Killing of George Floyd.

\begin{verbatim}
    {
        "date": "2020-05-25T00:00:00Z",
        "q": "black lives matter"
    }
\end{verbatim}

\paragraph{Brexit}

Focal date: Day of the referendum.

\begin{verbatim}
    {
        "date": "2016-06-23T00:00:00Z"
        "q": "brexit referendum"
    }
\end{verbatim}

\paragraph{Capitol riots}

Focal date: January 6th attack on the US Capitol.

\begin{verbatim}
    {
        "date": "2021-01-06T00:00:00Z",
        "q": "us capitol"
    }
\end{verbatim}

\paragraph{Grammys 2024}

Focal date: Day of the Awards ceremony.

\begin{verbatim}
    {
        "date": "2024-02-04T00:00:00Z",
        "q": "grammy awards"
    }
\end{verbatim}

\paragraph{Higgs Boson}

Focal date: Announcement of the ``God particle'' discovery.

\begin{verbatim}
    {
        "date": "2012-07-04T00:00:00Z",
        "q": "higgs boson"
    }
\end{verbatim}

\paragraph{World Cup 2014}

Focal date: Start and first game of the tournament.

\begin{verbatim}
    {
        "date": "2014-06-12T00:00:00Z",
        "q": "fifa world cup"
    }
\end{verbatim}

\section{ID-Based Queries}\label{app:id_endpoints}

This section covers tests conducted with API endpoints that accept video or other IDs as queries.
These endpoints show stable behavior and return mostly consistent data.

\subsection{Video: list Endpoint}

\begin{figure*}[t!]
  \centering
    \includegraphics[width=0.85\textwidth]{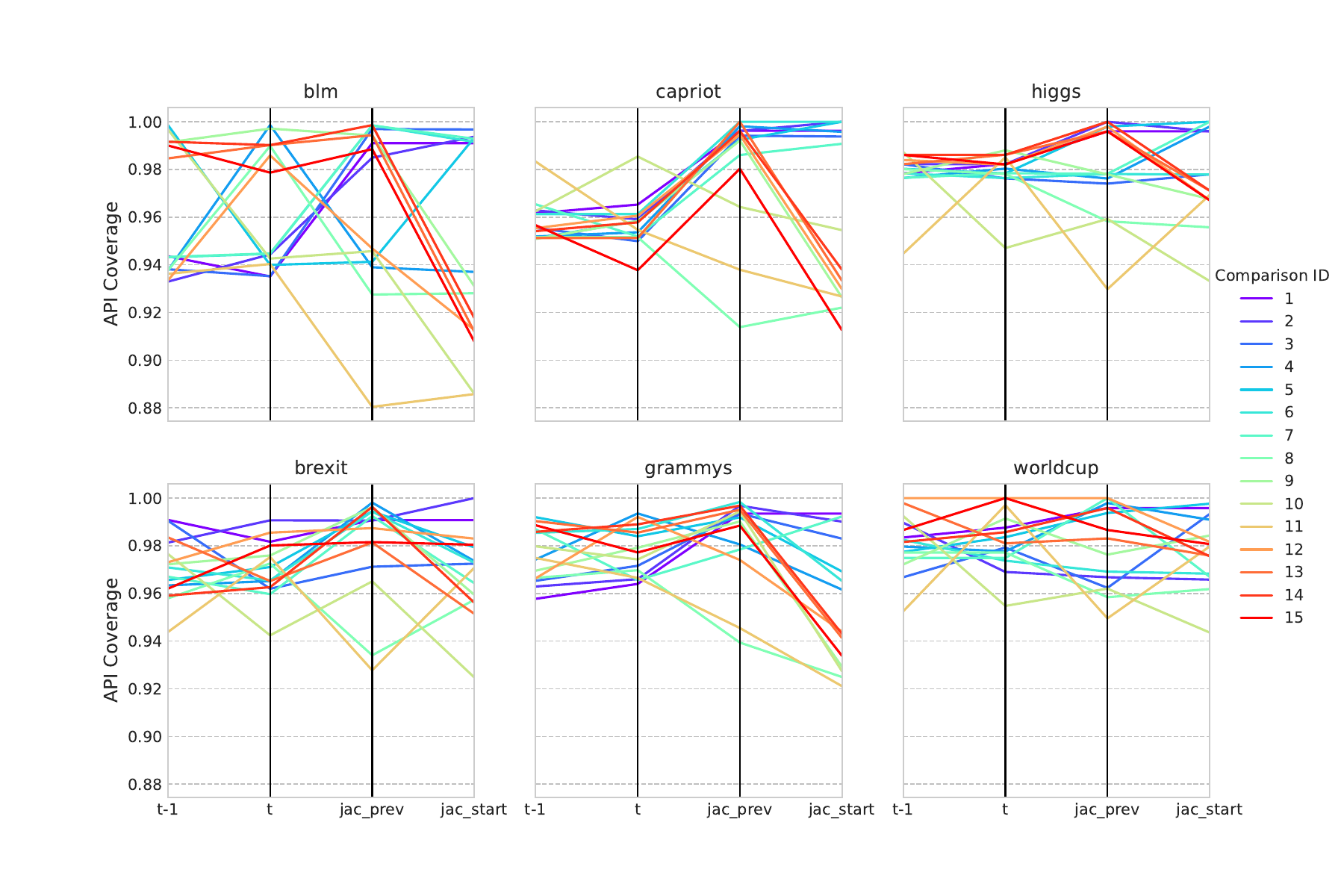}
    \caption{Parallel plots of percentage of common videos retrieved between successive runs and overall Jaccard similarity of common videos retrieved using the ``Videos: list'' endpoint. Comparison ID refers to the respective collection (higher ID = later collection).}
    \label{fig:dets_parallel}
\end{figure*}

In Figure~\ref{fig:dets_parallel}, we show common video IDs between collections at a given time \textit{t} and the previous time, as well as the first collection.
For each collection, we query the Video: list endpoint immediately after obtaining results through the search endpoint to get details and metadata about videos, such as their descriptions, view counts, likes, etc., using the video IDs.
We compute the percentage of videos for which metadata is returned at \textit{t} and \textit{t-1},
% (noting that these collections are performed immediately after the respective ``search'' collection), 
and we also obtain Jaccard similarities for the videos returned between $S_t$ and $S_{t-1}$, as well as between $S_t$ and $S_1$. 
Since these comparisons are restricted only to video IDs that are common in both sets being compared, the overall coverage and Jaccard similarity are higher for this endpoint.
Moreover, the fact that we do not find consistent patterns between comparison ID and $J(S_t,S_1)$ suggests that API gaps in returning specific video metadata are not systematic, and are thus likely errors rather than intentional API behavior.

\subsection{CommentThreads: list and Comments: list Endpoints}

\begin{table}[t!]
    \small
    \centering
    \begin{tabular}{lrrrrrr}
    \toprule
        \textbf{topic} & \textbf{TL, NS} & \textbf{N, NS} & \textbf{TL, S} & \textbf{N, S}  \\
        \midrule
        BLM & .329 & .307 & .976 & .983  \\
        Brexit & .381 & .339 & .999 & .999 \\
        Capitol & .648 & .625 & .998 & .994  \\
        Grammys & .728 & .737 & .996 & .992 \\
        Higgs & .974 & N/A & .998 & N/A \\
        World Cup & .470 & .532 & .999 & .999 \\
        \bottomrule
    \end{tabular}
    \caption{Jaccard similarities between first- and last-collection comment sets. TL = top-level, N = nested. NS = non-shared videos (full sets), S = shared videos only. N/A values for the Higgs topic, which is the oldest, are possibly due to the comment reply affordance operating differently in 2012.}
    \label{tab:comments}
\end{table}

We also query the CommentThreads: list endpoint, which accepts video IDs as queries and returns all comment threads (with a maximum of five nested comments), as well as the Comments: list endpoint, which accepts thread IDs as queries and returns all nested comments.
Due to the scale and number of comments and comment threads returned in each instance, we only make comparisons between the first and last collection.
Moreover, we only consider comments that were posted at most 3 weeks after the given topic's D-day (we allow an additional week beyond our video collection stopping point to allow for consolidation of comments on videos that were uploaded later).
Given the findings presented in Figure~\ref{fig:vid_jac}, we assume that any differences will be maximized the farther apart the collections occur.

For each collection instance, we compare the similarities of the set of top-level and nested comments returned across both all videos returned in that respective collection, and across videos that are common in both collections.
We show these results in Table~\ref{tab:comments}.
Unsurprisingly, we find some deviations between both top-level and nested comment sets between the first and last collections, as these are drawn from different parent videos.
However, the deviation patterns are not necessarily consistent with the magnitude of deviations in video IDs themselves.
For example, although the Brexit topic shows overall lower video deviations than other topics (except for Higgs), it shows the second-highest deviation behind only BLM in both top-level and nested comments; this may be an artifact of higher activity under more contested topics.
With respect to comments drawn from common videos, differences between both nested and top-level comments between collections are negligible, showing that this endpoint itself does not systematically randomize results and is likely returning (almost) all comments for every queried video.

% Given that the number of comments across videos ranges from tens of thousands (for Higgs) to hundreds of thousands (for all other topics), these deviations translate to several thousand comments of discrepancies between sets returned on different dates.
% Thus, merely the date on which a query is made may influence results.

\section{Regression Robustness Checks}\label{app:regressions}

Alternative regression model setups as robustness checks against the SHAP-explained gradient boosting model implemented in the main paper.

\subsection{Binned Ordinal Regression}

We split the frequency of returns into four roughly equal bins (1-5, 6-10, 11-15, 16), taking into account that 16 is the modal value.
% which results in four roughly equal bins (16 is the modal value).
We then perform an ordinal regression using a logit link function.
This function is chosen due to the uniform distribution arising from the binning.
All continuous features are log-transformed to reduce multicollinearity and standardized for better comparison between coefficients.

A log-likelihood test shows that this model significantly outperforms the null model ($\chi^2 = 1137.63, p < 0.001)$, although the overall fit is low (pseudo-$R^2 = 0.079$).
This suggests either that several other factors may influence video appearance, or that much of the variance is indeed random.
We show the standardized beta coefficients and confidence measures in Table~\ref{tab:regression_main}.
Topics are compared against BLM as the reference category, and the standard quality is compared against HD.

\begin{table}[t!]
    \small
    \centering
    \begin{tabular}{lrrr}
    \toprule
    \textbf{Variable} & $\beta$ & \textit{\textbf{SE}} & 95\% CI \\
    \midrule
        SD (quality) & -0.018 & 0.079 & [-0.173, 0.137] \\
        brexit (topic) & \text{***}1.231 & 0.098 & [1.039, 1.423] \\
        capriot (topic) & -0.160 & 0.093 & [-0.341, 0.022] \\
        grammys (topic) & *0.171 & 0.083 & [0.008, 0.333] \\
        higgs (topic) & \text{***}3.10 & 0.141 & [2.826, 3.379] \\
        worldcup (topic) & 0.161 & 0.101 & [-0.037, 0.359] \\
        duration & \text{***}-0.115 & 0.028 & [-0.170, -0.061] \\
        views & 0.161 & 0.088 & [-0.011, 0.333] \\
        likes & \text{**}0.285 & 0.095 & [0.098, 0.471] \\
        comments & 0.069 & 0.064 & [-0.058, 0.195] \\
        channel age & 0.049 & 0.031 & [-0.012, 0.110] \\
        channel views & \text{*}0.3176 & 0.135 & [0.053, 0.582] \\
        channel subs & \text{**}-0.3784 & 0.122 & [-0.617, -0.140] \\
        \#~channel videos & -0.0212 & 0.075 & [-0.169, 0.126] \\
        \bottomrule
    \end{tabular}
    \caption{Standardized regression coefficients for binned ordinal model. *$p < 0.05$, **$p < 0.01$, ***$p < 0.001$.}
    \label{tab:regression_main}
\end{table}

\subsection{Frequency as Continuous Variable}

\begin{table}[t!]
    \small
    \centering
    \begin{tabular}{lrrr}
    \toprule
    \textbf{Variable} & $\beta$ & \textit{\textbf{SE}} & 95\% CI \\
    \midrule
        SD (quality) & 0.0712 & 0.205 & [-0.331, 0.474] \\
        brexit (topic) & \text{***}3.416 & 0.274 & [2.878, 3.953] \\
        capriot (topic) & -0.283 & 0.257 & [-0.786, 0.220] \\
        grammys (topic) & \text{*}0.571 & 0.238 & [0.105, 1.038] \\
        higgs (topic) & \text{***}6.718 & 0.248 & [6.231, 7.205] \\
        worldcup (topic) & 0.438 & 0.288 & [-0.126, 1.003] \\
        duration & \text{***}-0.285 & 0.076 & [-0.435, -0.135] \\
        views & 0.429 & 0.238 & [-0.037, 0.896] \\
        likes & \text{**}0.713 & 0.262 & [0.198, 1.227] \\
        comments & 0.242 & 0.177 & [-0.105, 0.588] \\
        channel age & 0.113 & 0.084 & [-0.052, 0.279] \\
        channel views & \text{**}1.079 & 0.349 & [0.394, 1.763] \\
        channel subs & \text{***}-1.157 & 0.319 & [-1.783, -0.531] \\
        \#~channel videos & -0.2212 & 0.208 & [-0.629, 0.187] \\
        \bottomrule
    \end{tabular}
    \caption{Standardized regression coefficients for OLS model. $*p < 0.05$, $**p < 0.01$, $***p < 0.001$.}
    \label{tab:app_regcont}
\end{table}

We use frequency as our dependent variable in a multiple Ordinary Least Squares (OLS) regression with robust standard errors.
The overall model is significant ($F_{(14,5348)}=122.3, p < 0.001$) and shows modest fit ($R^2 = 0.164$).
We report standardized beta coefficients with confidence metrics in Table~\ref{tab:app_regcont}.
The patterns are identical to the previous binned ordinal regression model and to what is reported in the main paper.

\subsection{Non-Binned Ordinal Regression}

Ordinal regression where frequencies are treated as 16 distinct categories.
We use a complementary log-log link function instead of logit due to the distribution being skewed towards the highest value.
The overall model performs significantly better against a null model ($\chi^2 = 1167.64, P < 0.001$), although the overall fit is low (pseudo-$R^2 = 0.04$).
Coefficients are reported in Table~\ref{tab:app_regfullord}.
Patterns are largely consistent with the other models, except for the World Cup topic now also showing marginally significant differences (higher return frequencies) compared to BLM.

\begin{table}[t!]
    \small
    \centering
    \begin{tabular}{lrrr}
    \toprule
    \textbf{Variable} & $\beta$ & \textit{\textbf{SE}} & 95\% CI \\
    \midrule
        SD (quality) & 0.0228 & 0.051 & [-0.077, 0.122] \\
        brexit (topic) & \text{***}0.9207 & 0.065 & [0.793, 1.049] \\
        capriot (topic) & -0.0412 & 0.059 & [-0.156, 0.074] \\
        grammys (topic) & \text{***}0.2395 & 0.051 & [0.139, 0.340] \\
        higgs (topic) & \text{***}2.2998 & 0.115 & [2.075, 2.525] \\
        worldcup (topic) & \text{*}0.1338 & 0.066 & [0.004, 0.264] \\
        duration & \text{***}-0.0710 & 0.018 & [-0.106, -0.036] \\
        views & 0.0352 & 0.056 & [-0.074, 0.145] \\
        likes & \text{**}0.2051 & 0.062 & [0.084, 0.326] \\
        comments & 0.0656 & 0.042 & [-0.017, 0.148] \\
        channel age & 0.0355 & 0.019 & [-0.002, 0.073] \\
        channel views & \text{**}0.2852 & 0.093 & [0.103, 0.468] \\
        channel subs & \text{**}-0.2734 & 0.081 & [-0.431, -0.116] \\
        \#~channel videos & -0.0958 & 0.049 & [-0.193, 0.001] \\
        \bottomrule
    \end{tabular}
    \caption{Standardized regression coefficients for full ordered model. *$p < 0.05$, **$p < 0.01$, ***$p < 0.001$.}
    \label{tab:app_regfullord}
\end{table}

\end{document}